\documentclass[aps,prb,twocolumn,superscriptaddress,a4paper]{revtex4-2}
\pdfoutput=1 

\usepackage{amsmath}           
\usepackage{amsfonts}          
\usepackage{amssymb}           
\usepackage{graphicx}          
\usepackage[T1]{fontenc}       
\usepackage{subfigure}
\usepackage{url}
\usepackage{hyperref}
\hypersetup{colorlinks=true, citecolor=blue, urlcolor=blue, linkcolor=blue}
\usepackage[dvipsnames]{xcolor}
\newcommand{\EPSG}{EuPd$_2$(Si$_{1-x}$Ge$_x$)$_2$}
\newcommand{\EPS}{EuPd$_2$Si$_2$}
\newcommand{\EPG}{EuPd$_2$Ge$_2$}

\begin{document}
\title{From valence fluctuations to long-range magnetic order\\ in \EPSG\,single crystals} 
\author{Marius Peters}
\affiliation{Physikalisches Institut, Goethe-Universit\"{a}t Frankfurt, 60438 Frankfurt am Main, Germany}
\author{Kristin Kliemt}
\affiliation{Physikalisches Institut, Goethe-Universit\"{a}t Frankfurt, 60438 Frankfurt am Main, Germany}
\author{Michelle Ocker}
\affiliation{Physikalisches Institut, Goethe-Universit\"{a}t Frankfurt, 60438 Frankfurt am Main, Germany}
\author{Bernd Wolf}
\affiliation{Physikalisches Institut, Goethe-Universit\"{a}t Frankfurt, 60438 Frankfurt am Main, Germany}
\author{Pascal Puphal}
\affiliation{Max-Planck-Institute for Solid State Research, 70569 Stuttgart, Germany}
\author{Matthieu Le Tacon}
\affiliation{Institute for Quantum Materials and Technologies, Karlsruhe Institute of Technology, 76021 Karlsruhe, Germany}
\author{Michael Merz}
\affiliation{Institute for Quantum Materials and Technologies, Karlsruhe Institute of Technology, 76021 Karlsruhe, Germany}
\affiliation{Karlsruhe Nano Micro Facility (KNMFi), Karlsruhe Institute of Technology, 76344 Eggenstein-Leopoldshafen, Germany}
\author{Michael Lang}
\affiliation{Physikalisches Institut, Goethe-Universit\"{a}t Frankfurt, 60438 Frankfurt am Main, Germany}
\author{Cornelius Krellner}
\affiliation{Physikalisches Institut, Goethe-Universit\"{a}t Frankfurt, 60438 Frankfurt am Main, Germany}
\email{krellner@physik.uni-frankfurt.de}
\begin{abstract}                                          
\EPS\, is a valence-fluctuating system undergoing a temperature-induced valence crossover at $T'_V\approx160\,$K.
We present the successful single-crystal growth using the Czochralski method for the substitution series \EPSG, with substitution levels $x\leq 0.15$. A careful determination of the germanium content revealed that only half of the nominal concentration is built into the crystal structure. From thermodynamic measurements it is established that $T'_V$ is strongly suppressed for small substitution levels and antiferromagnetic order from stable divalent europium emerges for $x\gtrsim 0.10$. The valence transition is accompanied by a pronounced change of the lattice parameter $a$ of order 1.8\%. In the antiferromagnetically-ordered state below $T_N = 47$\,K, we find sizeable magnetic anisotropy with an easy plane perpendicular to the crystallographic $c$ direction. An entropy analysis revealed that no valence fluctuations are present for the magnetically-ordered materials. Combining the obtained thermodynamic and structural data, we construct a concentration-temperature phase diagram demonstrating a rather abrupt change from a valence-fluctuating to a magnetically-ordered state in \EPSG.
\end{abstract}

\maketitle

\section{Introduction}
Within the last years, attention has been drawn towards modeling the thermodynamic behavior of materials by explicitly considering a coupling between a material's electronic degrees of freedom and its lattice degrees of freedom. Propositions were made how to describe the entanglement between electronic thermodynamical and quantum phase transitions and the elastic responses of the crystal lattice in form of a (quantum) critical elasticity theory \cite{Garst15}. For example, when examining the Mott metal-insulator transition in the organic charge transfer salt $\kappa$-(BEDT-TTF)$_2$Cu$\left[\text{N(CN)}_2\right]$Cl, simultaneously to the electronic transition between the conducting and the insulating state, a breakdown of Hooke's law can be observed \cite{Gati18}, tying the behavior of the crystal lattice closely to the behavior of the electronic system. In some iron-based superconductors, a strong contraction of the $c$ direction of the crystal lattice can be observed, with strong effects on the magnetic and superconducting properties \cite{Kreyssig08, Soh13, Stillwell19}. More systems that offer electronic transitions accompanied by strong lattice effects shall be investigated in order to function as probe systems for the theoretical framework \cite{Garst15}.

By searching for such systems displaying closely linked electronic and lattice effects, europium-based intermetallic systems have shifted back into the focus of attention. In the 1980's, polycrystalline samples of \EPS\, served as a model system for valence fluctuations between two valence states of europium: Eu$^{2+}$ and Eu$^{3+}$
\cite{Sampathkumaran81a, Sampathkumaran81b}
being shiftable between different states of intermediate valence by manipulating the materials using 
temperature \cite{Kemly1985, Wortmann1985, AbdElmeguid1985, Holland1987, Mimura04} or external pressure \cite{Vijayakumar1981, Batlogg82,Srinivasan1984,Adams91}.
Large volume changes accompany the valence transition between the spatially larger Eu$^{2+}$  and the smaller Eu$^{3+}$  configuration, resulting in a shrinking of the \textit{a} lattice parameter of 0.18\,\AA\, when going from the Eu$^{(2+\delta)+}$ to the Eu$^{(3-\delta')+}$ valence state \cite{Sampathkumaran81a, Jhans87}. The precise europium valencies of \EPS\, were determined by hard X-ray photoelectron spectroscopy and values of Eu$^{2.23+}$ at 300\,K and Eu$^{2.75+}$ at 20\,K were determined \cite{Mimura11}.

Together with other europium-based intermetallic systems, displaying either valence-fluctuating states or magnetically-ordered Eu$^{2+}$  states, \EPS\ was located in a generalized $p-T$ phase diagram close to the critical endpoint of the valence transition based on investigations on single-crystalline samples \cite{Onuki17}. The Eu$^{2+}$  systems and their transitions into a long-range magnetically-ordered phase can be located at the low pressure side of the phase diagram. Towards higher pressures, two different intermediate valent states, the Eu$^{(2+\delta)+}$ state at high temperatures, and the Eu\textsuperscript{(3-$\delta$')+} state at low temperatures, occur. They are separated by a line of first order transitions at $T_V$, that ends in a critical endpoint of second order, beyond which a crossover area is entered at higher pressures. In this paper, we denote the temperature, at which this valence crossover occurs with $T'_V$. In a region in proximity to the critical endpoint, the changes in the electronic system of the europium valence might induce a critical elastic response in the crystal lattice. This makes EuPd\textsubscript{2}Si\textsubscript{2} a suitable target material for probing the predictions of critical elasticity theory.

Unsubstituted \EPS\, is already a promising candidate, displaying large lattice effects accompanying the change in valence
\cite{Sampathkumaran81a, Ye_2023}. Earlier investigations on polycrystalline material located the system on the high pressure side of the critical endpoint in the crossover area \cite{Batlogg82}. In order to use this system as a testbed for probing  anomalous behavior related to the proximity of a critical endpoint in pressure studies, \textit{negative} pressure would need to be exerted to shift the system to the low pressure side. Usually, this is done by substituting one of the elements of the compound partly by a larger element, forcing the unit cell to expand. For polycrystalline samples of EuPd\textsubscript{2}Si\textsubscript{2}, such substitutions have been performed, replacing palladium with platinum \cite{Wada01} and gold \cite{Segre82}, or silicon with germanium \cite{Cho02} and tin \cite{Mishra83}. Recently, we have shown that a change of the Pd-Si ratio in \EPS\, can also cause a shift in the valence crossover temperature \cite{Kliemt22}, which explains the different values of $T'_V$ reported in literature for this system \cite{Wada01, Mimura11, Onuki17}. In addition, it was possible to grow epitaxial thin films of \EPS\, on Mg(001) substrates using molecular beam epitaxy \cite{Koelsch22}. Due to a clamping effect of the \EPS\, thin film to the MgO substrate with negligible thermal expansion, the abrupt change of the lattice parameter $a$ of \EPS\, is suppressed, leading to a highly strained thin film upon cooling, which do not show a valence transition anymore, but probably a magnetically-ordered ground state \cite{Koelsch22}.

In this work, we will focus on the germanium-substituted system, \EPSG, bringing two new approaches to what is known about the system so far: First, we will shift attention towards the silicon rich, valence-fluctuating regime of the system, while previous works focused on its germanium rich, long-range antiferromagnetically-ordered regime \cite{Cho02}. Cho \textit{et al.} \cite{Cho02} were able to give an estimation of a substitution level $x_M\approx 0.15$ for polycrystalline samples, at which long-range magnetic order breaks down and is replaced by valence-fluctuating behavior, but no systematic characterization of this crossover region has been done. A more detailed investigation of electronic and lattice behavior is due in order to map the suppression of valence fluctuations and the occurrence of antiferromagnetism in the system. Second, we apply the Czochralski method to grow single-crystalline samples of germanium-substituted \EPS. These large crystals will allow for a proper characterization of magnetic anisotropies and elastic responses in the crystal lattice in proximity to the critical endpoint of the proposed first-order valence transition. The present study is complemented by detailed investigations of the effect of hydrostatic (He-gas) pressure on selected crystals \cite{Wolf23}.   

\section{Experimental Details}
Czochralski growth of \EPSG\, was performed for different nominal substitution levels $0.05\le x \le0.30$ after a procedure established in previous work \cite{Kliemt22}. High purity materials Eu (99.99\%, chunks, EvoChem), Pd (99.99\%, rod, Heraeus), Si (99.9999\%, pieces, Cerac) and Ge (99.9999\%, pieces, Otavi Minen) with an initial stoichiometry of Eu$_{1.45}$Pd$_2$(Si$_{1-x}$Ge$_x$)$_2$ and an initial mass of 15\,g were used. Before performing the actual Czochralski growth process, two steps of prereaction were applied to make the materials accessible in the growth experiment. To overcome high melting temperatures of palladium ($1555^\circ$C) and silicon ($1414^\circ$C), in a first step palladium, silicon and germanium were melted together by arc melting. Since both binary compounds PdSi and PdGe melt at about $900^\circ$C, a comparable lowering of the melting temperature can be expected in the given ternary case. In a second step, the prereacted Pd, Si and Ge are brought together with Eu (melting temperature $T_M=826^\circ$C) in a glassy carbon inner crucible, sealed inside a niobium outer crucible, and heated to $835^\circ$C for 1\,h under an argon protective atmosphere (box furnace by Linn company).

\begin{figure}[b]
	\begin{center}
		\subfigure{\includegraphics[width=0.95\linewidth]{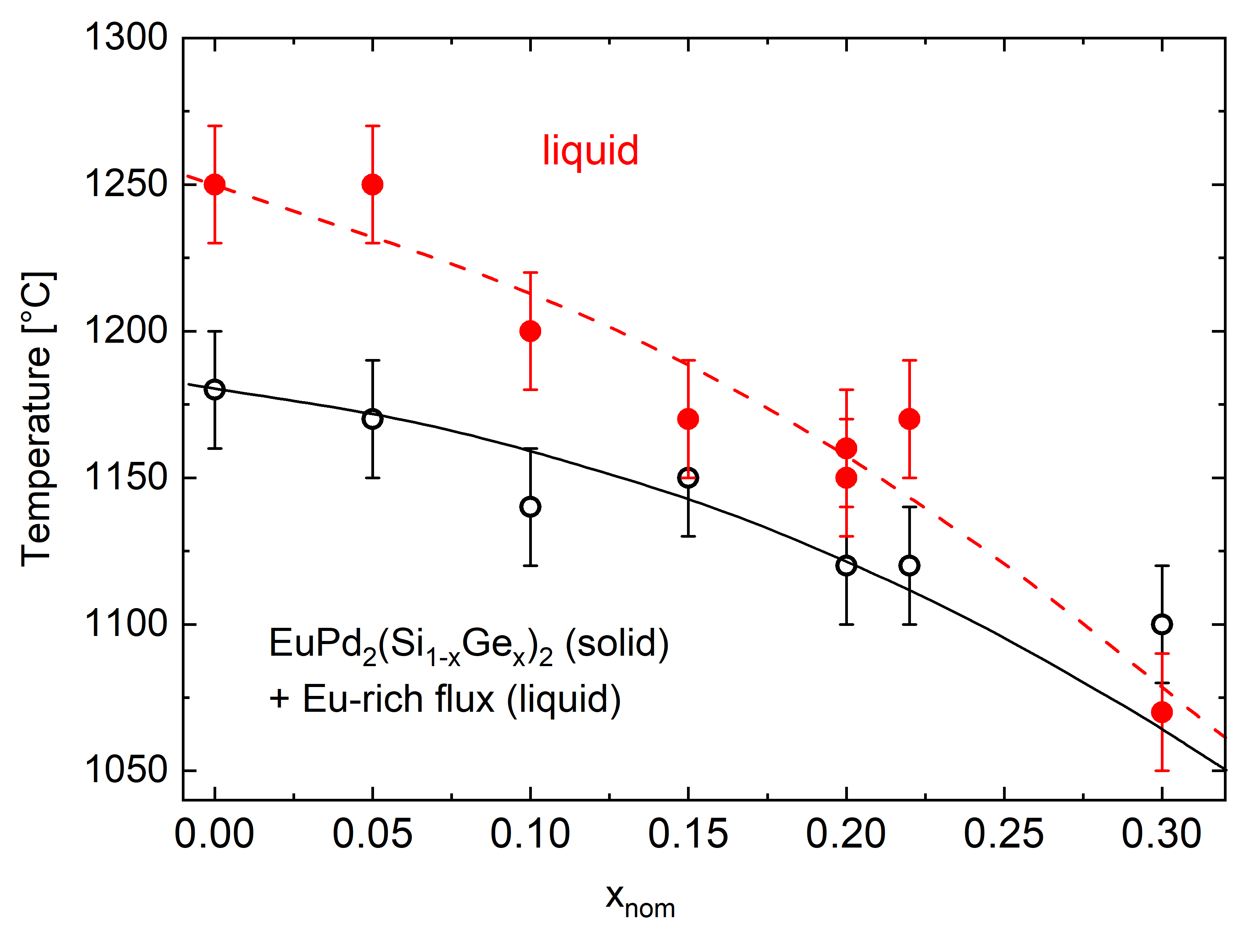}}
		\subfigure{\includegraphics[width=0.95\linewidth]{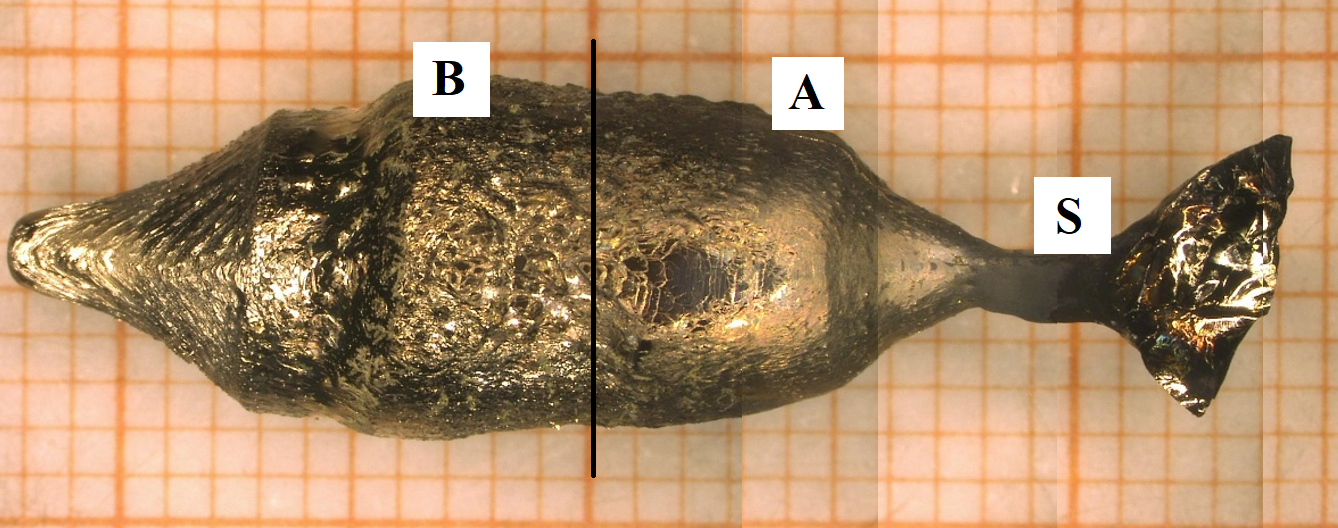}}
		\caption{Top: Melt composition - temperature phase diagram for the growth of \EPSG\, single crystals. Open black (full red) symbols mark the temperatures where the first signs of solidification (melting) are observable during cooling (heating) of the precursor of different nominal Ge content $x_{\text{nom}}$. Dashed and solid lines are guides to the eyes. Bottom: Result of the Czochralski growth process of the sample with nominal $x=0.10$. S marks the seeding crystal, A the area where the target phase crystallizes, and B the part where secondary phases regularly appears.}
		\label{sample}
	\end{center}
\end{figure}

The actual Czochralski growth is performed in a growth chamber by Arthur D. Little, the precursor is inductively heated with a H\"uttinger generator. For the growth process, argon over pressure of 20\,bar was applied to slow down europium evaporation from the melt. During the process, the melt levitates by virtue of an inhomogeneous magnetic field within a cold copper crucible. We used this method since previously strong reactions with any accessible crucible materials were seen. In the top part of Fig.~\ref{sample}, we show the liquidus temperatures as function of the nominal Ge-concentration, measured with an IRCON pyrometer during the different Czochralski growths. The area above the full red points denotes the homogeneous melt after heating. The temperature where solidification starts during cooling down, without a seed crystal, is marked as open black points. Therefore, the area between the red and black curves is the so-called Ostwald-Miers area, where crystallization without new nucleation is possible \cite{Scheel11}. This area is shifted to lower temperatures and gets narrower with increasing $x_{\text{nom}}$, which means that the occurrence of competing grains during the growth gets more likely with increasing $x_{\text{nom}}$.

Five substitution levels with nominal concentrations of $x_{\text{nom}}$ = 0.05, 0.10, 0.15, 0.20, 0.30 were prepared. As a reference for the overall characterization, results from the unsubstituted system \cite{Kliemt22} were included in the discussions. Czochralski growth experiments were performed at seeding temperatures between $1150^\circ$C and $1250^\circ$C (see Fig.~\ref{sample}), and the samples were pulled with rates between 0.8\,mm/h and 3.3\,mm/h. Growth experiments were seeded iteratively with single crystals from previously grown samples with neighboring $x$. The initial seed for the $x = 0.10$ sample stemmed from the previously grown unsubstituted \EPS\, system.

The samples were characterized using a variety of probing techniques. Sample composition and germanium incorporation were determined by Energy Dispersive X-ray analysis (EDX) using a Zeiss-DSM940A Scanning Electron Microscope with an EDAX detector. Lattice parameters were quantified by temperature-dependent powder X-ray diffractometry (PXRD) utilizing a Siemens D500 diffractometer with a helium-gas cooling system capable of reaching temperatures down to 10\,K and using Cu K$\alpha$ radiation. Refinement of the X-ray powder diffractometry data was performed using GSAS II \cite{GSASII}. Single crystal X-ray diffraction data on representative EuPd$_2$(Si$_{1-x}$Ge$_x$)$_2$ samples from the corresponding batches were collected at 295~K on a STOE imaging plate diffraction system (IPDS-2T) using Mo $K_{\alpha}$ radiation. For the investigated specimen all accessible reflections ($\approx 5100$) were measured up to a maximum angle of $2 \Theta =65^{\circ}$. The data were corrected for Lorentz, polarization, extinction, and absorption effects. Using SHELXL \cite{Sheldrick08} and JANA2006 \cite{Petricek14}, all averaged symmetry-independent reflections ($I > 2 \sigma$) were included for the refinements. For all compositions the unit cell and the space group were determined, the atoms were localized in the unit cell utilizing random phases as well as Patterson superposition methods, the structure was completed and solved using difference Fourier analysis, and finally the structure was refined. In all cases the refinements converged quite well and show excellent reliability factors (see GOF, $R_1$, and $wR_2$ in Table \ref{tab:SCXRD}). 

Sample orientation was carried out using a Laue camera with white X-ray radiation from a tungsten anode. Heat capacity, resistivity, and magnetization  of the samples were measured using the standard measurement options (HC, ACT, VSM) of a 9\,T Quantum Design PPMS. 
Some measurements of the magnetic susceptibility were performed by utilizing a commercial superconducting quantum interference device (SQUID) magnetometer (MPMS, Quantum Design). The ac-susceptibility was measured at different frequencies with the corresponding option of a MPMS3 using a driving field of 4~Oe.

\begin{table}[htbp]
	\centering
	\begin{tabular*}{8.5cm}{@{\extracolsep{\fill}}|c|c|c|c|c|c|}
		\hline
		\hline
		\rule[-7pt]{0pt}{20pt}  $x_{\text{nom}}$  & $x_{\text{EDX}}$ & $a$ [\AA] &  $c$ [\AA] & $T'_{V}$ [K] & $T_{N}$ [K]\\
		\hline
		\rule[-6pt]{0pt}{20pt}   0  & 0 & 4.2392(6) & 9.8674(12) & 140-160 & - \\
		\rule[-6pt]{0pt}{20pt}   0.05 & 0.034(6) & 4.2507(6)& 9.8704(20) & 105 & - \\
		\rule[-6pt]{0pt}{20pt}   0.10 & 0.058(7) & 4.2549(5)& 9.8722(15) & 87 & -  \\
		\rule[-6pt]{0pt}{20pt}   0.15 & 0.089(11) & 4.2584(3) & 9.8940(11) & 54/64 & -  \\
		\rule[-6pt]{0pt}{20pt}   0.20 & 0.105(8) & 4.2828(6)& 9.9047(18) & - & 47  \\
		\rule[-6pt]{0pt}{20pt}   0.30 & 0.154(9) & -&-  & - & 42 \\
		\rule[-6pt]{0pt}{20pt}   1    & 1 & 4.376 & 10.072 & - & 17 \\
		\hline
		\hline
	\end{tabular*}  
	\caption[]{Germanium concentration of the different \EPSG\, crystals investigated, together with lattice parameters $a$ and $c$ at 295 K from single-crystal XRD. 
 The characteristic temperatures of either valence fluctuations ($T'_V$) or long-range antiferromagnetic order ($T_N$) are determined from heat capacity and magnetic susceptibility data. The data for $x=0$ and $x=1$ are from Ref. \cite{Kliemt22} and \cite{Onuki20}, respectively.
}
	\label{GEincorp}
\end{table}

\begin{table*}[t]
\caption{\label{tab:SCXRD} Crystallographic data for EuPd$_2$(Si$_{1-x}$Ge$_x$)$_2$ at 295 K determined from single-crystal X-ray diffraction. The nominal value of the germanium content $x_{\text{nom}}$ is given together with the values derived from the detailed structural analysis, denoted by $x_{\text{XRD}}$. The structure was refined in the tetragonal space group (SG) $I4/mmm$ for which the lattice parameters $a$ and $c$  are shown together with the volume $V$ of the unit cell. Eu sits on a $2a$ Wyckoff position with coordinates $0, 0, 0$, Pd on a $4d$ position with coordinates $\frac{1}{2}, 0, \frac{1}{4}$, and (Si,Ge) on a $4e$ position with coordinates $0, 0, z$.\@ The $U_{ii}$ denote the anisotropic atomic displacement parameters (ADPs) (for all these special positions $U_{11}=U_{22}$ and $U_{12}=U_{13}=U_{23}=0$). For all samples a certain amount $m$ of Si (or Ge) is found on the Pd site (Wyckoff position $4d$). For completeness the bond distances for Eu-(Si,Ge), Eu-Pd, Pd-(Si,Ge), and (Si,Ge)-(Si,Ge) along the crystallographic $c$ direction are depicted as well. The refinement for $x = 0$ is reproduced from Ref. \cite{Kliemt22}.  Errors shown are statistical errors from the refinement.}
\begin{ruledtabular}
		\begin{tabular}[t]{lccccc} 
                $x_{\text{nom}}$      &              0          &          0.05       &      0.10     &   0.15 & 0.20  \\
                $x_{\text{XRD}}$      &              0          &          0.034(9)       &       0.070(8)     &   0.111(7) & 0.174(11)  \\ \hline
     &  \multicolumn{5}{c}{lattice parameters $a$, $c$}  \\
$a$ (\AA)              &    4.2392(6)       &     4.2507(6)    &       4.2549(5)              &   4.2584(3) & 4.2828(6)  \\
			  $c$ (\AA)      &        9.8674(12)     &        9.8704(20)   &       9.8722(15)  & 9.8940(11) & 9.9047(18) \\
			  $V$ (\AA$^3$)            &      177.3           &     178.3       &      178.7  & 179.4       & 181.7 \\ 
     &  \multicolumn{5}{c}{Eu, $2a$: $0, 0, 0$ }  \\
			 $U_{\rm 11}$ (\AA$^2$) &   0.00866(27)     &      0.00906(23)   &     0.00945(24) & 0.00780(25) & 0.01168(73) \\ 
             $U_{\rm 33}$ (\AA$^2$) &   0.00961(31)     &      0.00668(27)   &     0.01066(27) & 0.00719(29) & 0.01093(91) \\ 
   & \multicolumn{5}{c}{Pd, $4d$: $\frac{1}{2}, 0, \frac{1}{4}$}  \\
			 $U_{\rm 11}$ (\AA$^2$) &   0.00985(25)     &      0.01035(26)   &     0.01151(24) & 0.00979(24) & 0.01497(72) \\ 
             $U_{\rm 33}$ (\AA$^2$) &   0.00965(29)     &      0.00689(30)   &     0.01131(28) & 0.00722(30) & 0.01248(105) \\ 
             $m$ (\%) &   2.9(5)     &      3.2(6)   &     2.2(4) & 2.7(3) & 3.8(12)  \\ 
    &       \multicolumn{5}{c}{(Si,Ge), $4e$: $0, 0, z$}  \\
			 $z$                 &    0.37783(19)        &      0.37738(9)    &      0.37718(17)  & 0.37697(17) & 0.37787(53) \\
			 $U_{\rm 11}$ (\AA$^2$) &   0.00944(58)     &      0.00997(83)   &     0.01134(71) & 0.01038(64)  &  0.01171(180) \\ 
             $U_{\rm 33}$ (\AA$^2$) &   0.01229(85)     &      0.01010(99)   &     0.01490(86) & 0.01262(79)  &  0.01151(247) \\ 
     &      \multicolumn{5}{c}{selected bond lengths}  \\
			 Eu-(Si,Ge) (\AA)         &    3.2309(9)         &   3.2402(10)    &    3.2438(8) & 3.2479(7)  & 3.2611(19) \\			
			 Eu-Pd (\AA)             &    3.2524(7)         &   3.2567(9)    &    3.2584(7) & 3.2637(5)    & 3.2737(9) \\				
			 Pd-(Si,Ge) (\AA)         &    2.4665(11)         &   2.4694(12)    &    2.4703(10) & 2.4722(9) & 2.4883(31) \\			
			 (Si,Ge)-(Si,Ge) (\AA)     &    2.4110(30)         &   2.4211(31)    &     2.4252(29) & 2.4341(22) &  2.4194(71) \\			
      &     \multicolumn{5}{c}{Goodness of fit and R values}  \\
			 GOF                       &       1.90           &         1.69        &       1.55 &  1.72  & 2.32 \\
			 $wR_2$ (\%)               &       4.15           &         3.91        &       3.53 & 3.85   & 7.29 \\
			 $R_1$ (\%)                &       1.72           &         1.40        &       1.45  & 1.63  & 3.57 \\ 
			 \end{tabular}
\end{ruledtabular}
\end{table*}

\section{Results and Discussion}
\subsection{Crystal growth and germanium incorporation}
The resulting crystals for each crystal growth experiment look similar to the one shown at the bottom of Fig.~\ref{sample}. In this figure, \textbf{S} marks the seeding crystal, which for the nominal $x=0.10$ growth stemmed from unsubstituted \EPS.
\textbf{A} denotes the area, in which the target phase could usually be found without or with a minor amount of inclusions of a secondary phase. The mass of part \textbf{A} is  between 3\,g and 5\,g. 
Here, facets were regularly found to ease the first orientation of the sample. \textbf{B} marks the area in which an Eu-rich secondary phase occurred more regularly, stemming from the Eu excess in the original melt. Therefore, the growth direction in Fig.~\ref{sample} is from right to left. The germanium concentration $x_{\text{EDX}}$ in the crystal was determined using the EDX method. The results of this analysis concerning the germanium distribution are shown in Table \ref{GEincorp} for the different nominal stoichiometries. The germanium incorporation rate is between 50\% and 70\%, being higher for lower germanium concentration in the melt. This value was determined by performing between 30 and 60 single-point EDX analyses over the whole length of section \textbf{A} of the respective crystal, and then determining statistical mean and standard deviation assuming a Gauss distribution. We note that the germanium content determined by single-crystal XRD, $x_{\text{XRD}}$, is higher for the nominal 0.15 and 0.20 batches (see Table~\ref{tab:SCXRD}). Nevertheless, we will use the EDX content as $x$-value when discussing the physical characterization measurements, as the measured samples were characterized by EDX. 

For small $x$, the germanium concentration is constant over the whole length of the crystal. For $x_{\text{nom}}\geq 0.20$, also in the \textbf{A} region of the crystal two coexisting phases were observed, one being the target phase, and another quaternary phase with a 
higher europium content and a silicon/germanium ratio of about 3:1. The amount of this secondary phase  varies along the growth direction for the different growth batches and is clearly visible as a secondary phase coexisting with the target phase in the electron microscope images as well as through additional peaks in the PXRD data. For the physical measurements the crystals were chosen such, that the contribution from the additional phase is as small as possible. However, for the $x_{\text{nom}}=0.30$ growth, this phase gets rather dominant and it was not possible to extract a phase-pure crystal with the \EPSG\, target phase.\\

\subsection{Magnetic susceptibility data}
\begin{figure}[tb]                                 
	\begin{center}                                           
		\includegraphics[width=\linewidth]{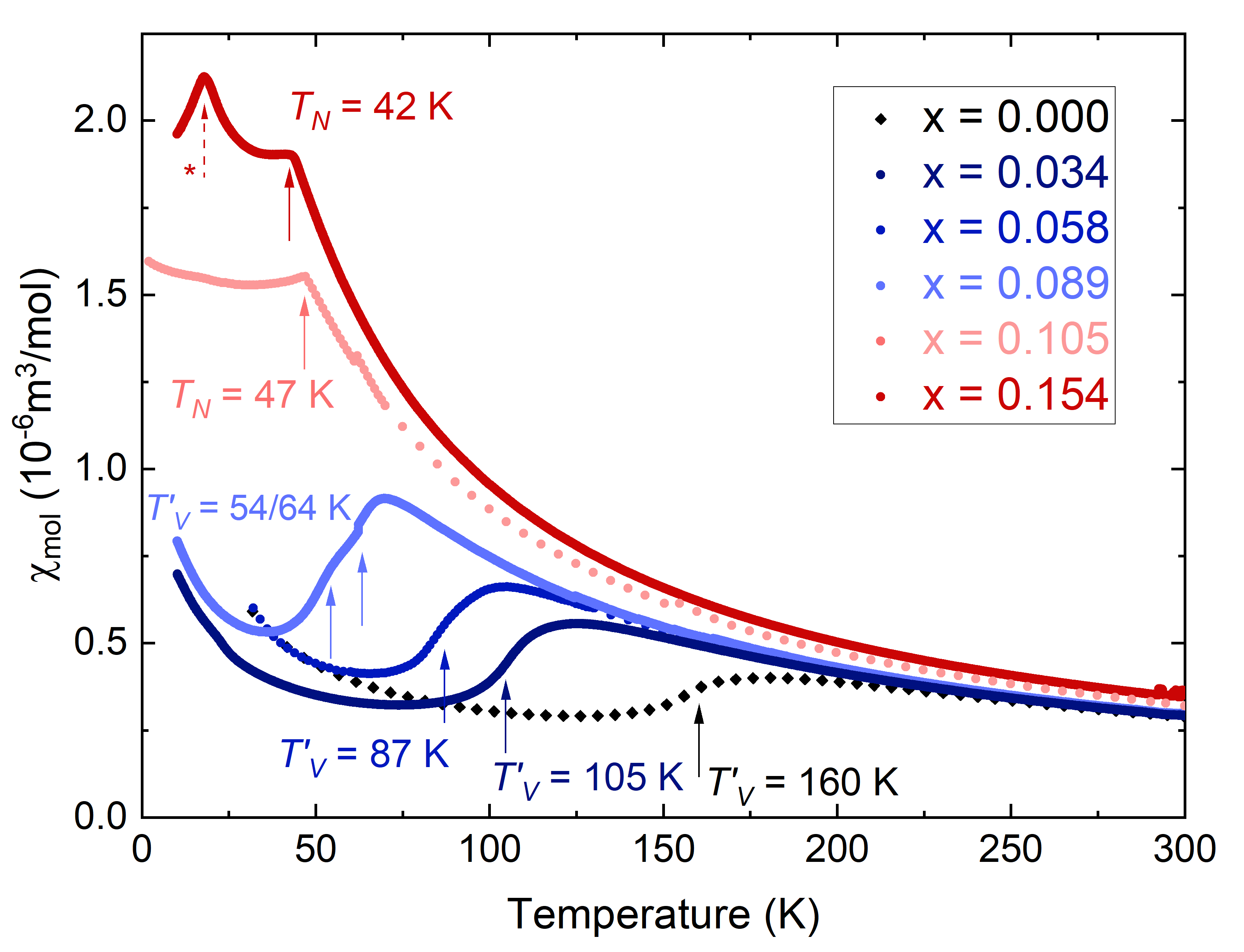} 
		\caption[]{Magnetic susceptibility data for different substituted samples at $B = 0.1$\,T and $B\,||\,c$. Black and blue arrows mark the valence crossover at $T'_V$, while red arrows indicate the AFM transition at $T_N$. The dashed arrow with a star denotes a magnetic transition from a secondary phase in the sample with $x=0.154$.}  
		\label{VSM}                                    
	\end{center}
\end{figure}

In Fig.~\ref{VSM}, we present the data of the magnetic susceptibility as a function of temperature for all different germanium concentrations investigated. A drastic change in the overall temperature dependence is observed between the samples with $x = 0.089$ (light blue) and $x = 0.105$ (light red). As we will later show in detail, the ground state of the two crystals is markedly different, although their Ge-concentration varies by only $\Delta x = 0.016$. In order to locate the temperature of the valence crossover from the magnetic susceptibility data for $x<0.10$, the maximum of the quantity d$(\chi(T)\cdot T)/$d$T$ was used. This quantity is proportional to the magnetic contribution to the heat capacity, and for the valence-fluctuating systems, its maximum corresponds to the inflection point of the magnetic susceptibility, see also Ref.~\cite{Wolf23} for details. For the antiferromagnetic (AFM) samples (red data), the N\'{e}el temperature was identified with the position of the sharp kink in $\chi(T)$, below which distinct magnetic anisotropy develops. Table \ref{GEincorp} gives an overview over the characteristic temperatures (valence crossover and magnetic transition) extracted from the susceptibility data.

All samples with $x\ge0.10$ also show a second anomaly at 17\,K, most clearly visible for $x=0.154$. Since the prominence of this anomaly is strongly sample dependent, this hints towards a secondary phase that is included in the sample. As discussed above, we have observed a quaternary compound in EDX  as secondary phase for the higher Ge concentrations. For $x_{\text{nom}}\geq 0.20$ this phase gets dominant in the B section of the grown crystal (see Fig.~\ref{sample}) and we were able to extract this unknown phase 
for a susceptibility measurement (not shown). This revealed the characteristics of a typical antiferromagnetically-ordered Eu$^{2+}$ system, with $T_N=17\,$K and a Curie-Weiss behavior above 50\,K with an effective moment of 7.9\,$\mu_B$ per europium. Therefore, we can attribute the second anomaly in the susceptibility data of Fig.~\ref{VSM} at 17\,K (dashed arrow with a star) to this additional phase.

\begin{figure}[t]                                 
	\begin{center}                                      
		\includegraphics[width=\linewidth]{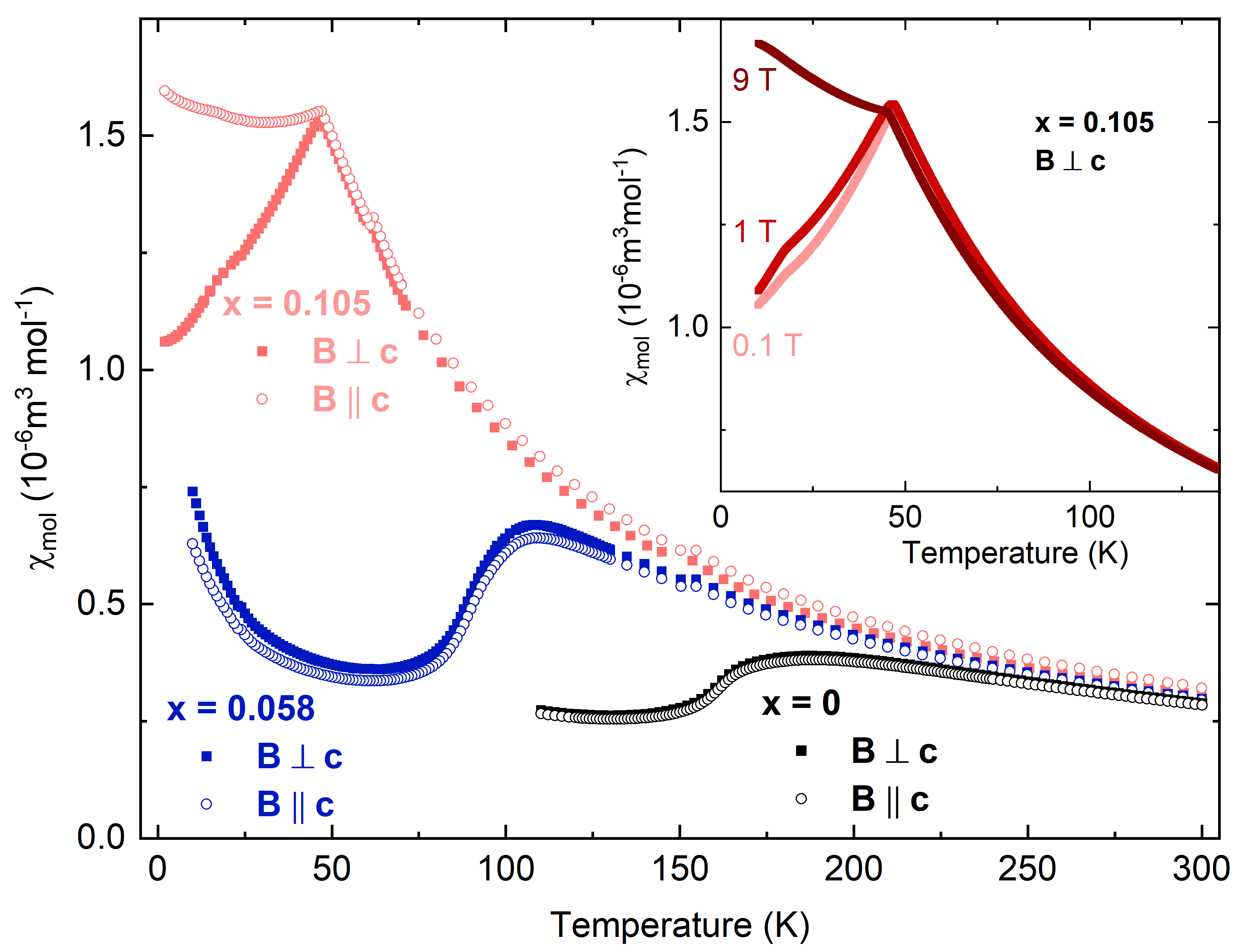}
		\caption[]{Magnetic susceptibility data at 1\,T for magnetic field parallel (open symbols) and perpendicular (closed symbols) to $c$ for different germanium substitutions of $x=0.105$ (red) and $x=0.058$ (blue) in comparison to the $x=0$ data (black). The $x=0.105$ sample presents clear magnetic anisotropy below $T_N$, while the $x=0.058$ and $x=0$ sample show no pronounced magnetic anisotropy. In the inset the magnetic-field dependence of the susceptibility for $B\perp c$ in the ordered state is shown for the $x=0.105$ crystal.  }  
		\label{AFM}                                    
	\end{center}
\end{figure}

\begin{figure}[t]                                 
	\begin{center}                                      
		\includegraphics[width=\linewidth]{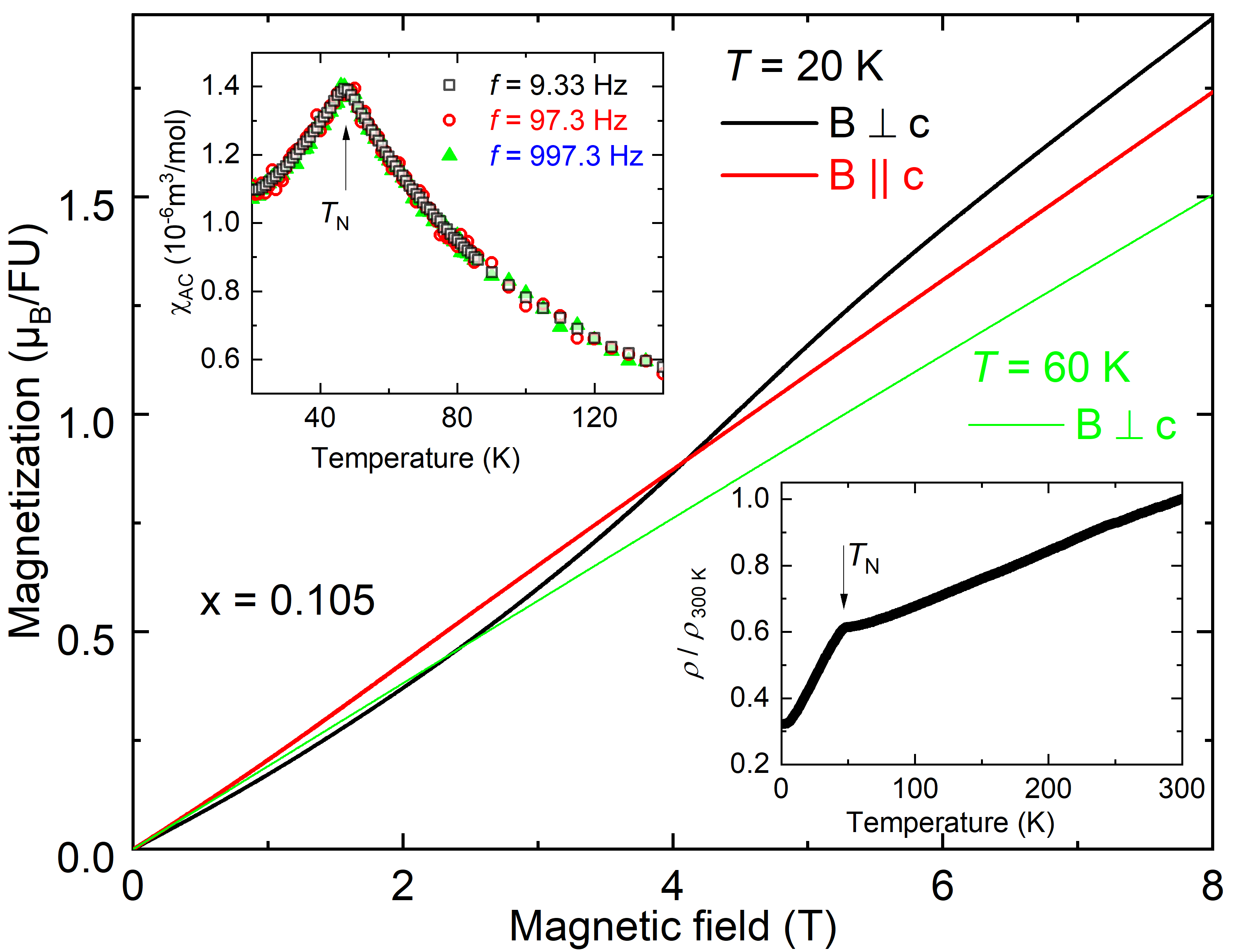}
		\caption[]{Characteristics of the AFM phase appearing in the  $x=0.105$ sample. Magnetization as function of the magnetic field below $T_N$ for magnetic field perpendicular (black) and parallel (red) to the $c$ direction and above $T_N$ (green). The upper left inset shows the ac-susceptibility measured with a driving field of 4~Oe perpendicular to $c$ at different frequencies. The lower right inset presents resistivity as function of temperature in zero field with the current applied perpendicular to the $c$ direction, showing a sharp anomaly at $T_N$.  }
		\label{fig_AFM}                                    
	\end{center}
\end{figure}

Comparative studies of the magnetic susceptibility for two different directions of the magnetic field ($B\perp c$ and $B \parallel c$) are presented in Fig.~\ref{AFM}. The difference between the samples with $x\le0.089$ and $x\ge0.105$ becomes again very apparent. Samples with $x\ge0.105$ show pronounced magnetic anisotropy below the transition temperature, with the magnetic easy plane perpendicular to the crystallographic $c$ axis. This, along with other characteristics discussed below, identifies them clearly as antiferromagnets with a N\'{e}el temperature of 47\,K and 42\,K for $x=0.105$ and $x=0.154$, respectively. Samples with $x\le0.089$ on the other hand do not show significant magnetic anisotropy below the transition. In addition, the small anisotropy at high temperatures is reversed for the two cases: Whereas the magnetically-ordered systems show an easy-plane anisotropy, the valence-fluctuating crystals are characterized by an almost isotropic magnetic response with a slightly lower out-of-plane susceptibility. In the inset of Fig.~\ref{AFM}, we present the field dependence of the susceptibility for in-plane magnetic fields up to 9\,T. The strong decrease below $T_N$ is observed for fields below 1\,T. Only the hump around 17\,K, due to the magnetic secondary phase discussed earlier, is affecting this. For 9\,T, the magnetic-ordering temperature is slightly suppressed down to 45~K, which is in agreement with AFM order. Below $T_N$, the temperature dependence is strongly affected by the higher magnetic field and a pronounced increase is apparent at 9\,T. For the other field direction, $B\, ||\, c$, the field dependence of the susceptibility is considerably weaker (not shown). Also this behavior is in agreement with AFM ordering of Eu$^{2+}$ moments with the magnetic easy plane perpendicular to the $c$ direction. For the crystals with $x\le0.089$, we do not observe a pronounced field dependence of the susceptibility and no comparable anisotropy below $T'_V$ for fields up to 9\,T.\\

To further characterize the AFM ordered phase in the $x=0.105$ sample, we present magnetization measurements at 20~K in the main part of Fig.~\ref{fig_AFM}. There, a weak metamagnetic transition around 4.5~T is observed for a magnetic  field perpendicular to $c$, whereas the curve for field along $c$ is a straight line. Above $T_N$ at 60~K, the curve with the in-plane field is again linear in $B$, as expected for a paramagnetic system. This overall behavior is in agreement with the above discussed susceptibility data, which classifies this material as an easy-plane antiferromagnet below $T_N$. In addition, we show ac-susceptibility data with a small driving field of 4~Oe perpendicular to $c$ in the upper left inset of Fig.~\ref{fig_AFM}. The motivation behind was to exclude signatures of spin-glass behavior in this material, which might occur due to the statistical substitution of Ge on the Si site. However, we do not observe any frequency dependence in the range of $f$ from $10-1000$~Hz, excluding spin-glass behavior below $T_N$. Accordingly, we do not observe any differences in the susceptibility in field-cooled versus zero-field-curve measurement protocols.\\
Finally, we present resistivity measurements (with current $\perp c$) for the $x=0.105$ sample in the lower right inset of Fig.~\ref{fig_AFM}. The temperature dependence above $T_N$ is linear in $T$ with a pronounced drop at $T_N$. This is a typical behavior for an Eu-based system with localized moments undergoing magnetic order, as e.g. EuGe$_2$Si$_2$ \cite{Onuki17} or EuRh$_2$Si$_2$ \cite{Seiro14}. However, this temperature dependence is in contrast to the behavior of a system with a valence crossover, which usually results in a broad upturn above the crossover temperature as observed in \EPS\,\cite{Onuki17, Kliemt22}. Therefore, also the resistivity data is in agreement with the proposed scenario, that the $x=0.105$ sample is an ordered AFM system without any signatures of a valence crossover.

\begin{figure}[t]                                 
	\begin{center}                                      
		\includegraphics[width=\linewidth]{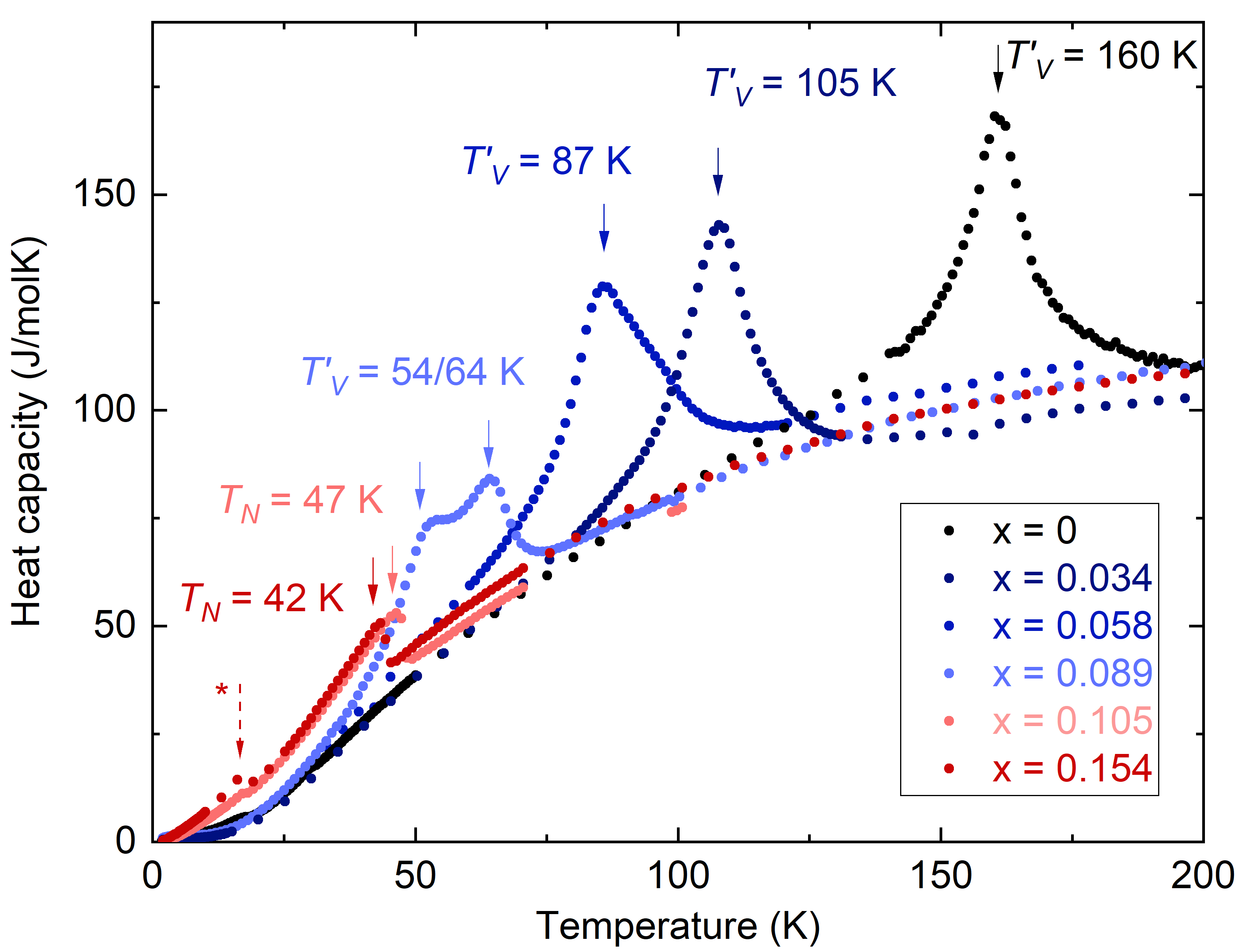}
		\caption[]{Heat capacity for single-crystalline \EPSG\, for varying $x$ as specified in the figure. Blue arrows mark the position of the characteristic temperature of a valence crossover, whereas red arrows mark the N\'{e}el temperature of the AFM transition. The dashed arrow with a star denotes an anomaly, assigned to the impurity phase, also seen in the susceptibility data.}  
		\label{HC}                                    
	\end{center}
\end{figure}

\subsection{Heat-capacity measurements}
Further evidence for the different nature of the observed phase transitions comes from heat-capacity measurements between 2 and 200 K shown in Fig.~\ref{HC}. The shape of the heat-capacity anomalies differs strongly between the samples with $x\le0.089$ (blue, symmetrical) and $x\ge0.105$ (red, mean-field type), showing the difference between valence-fluctuating and AFM samples. The double-peak anomaly in the $x=0.089$ sample is probably due to a small phase separation between two areas with slightly different substitution level. Within the accuracy of our EDX measurements this separation could not be resolved, so it is of the order of the error bar of the EDX measurements given in Table~\ref{GEincorp}. A slight shoulder of similar origin is probably seen for the $x=0.058$ data. At 17\,K, we observe a small hump in the heat-capacity data for the samples with $x\ge0.10$, which is more pronounced for the $x=0.154$ sample. Also here, we observe varying significance from sample to sample and relate this to the magnetic impurity phase, similar to what was observed in the susceptibility data. Having in mind the generally established phase diagram of Eu-based valence-fluctuating systems under pressure \cite{Onuki17}, we carefully measured large heating pulses ($\Delta T \approx 15$\,K) covering the transition at $T_N$ or $T'_V$, following the procedure described in Ref.~\cite{Lashley03}, but none of the samples showed latent heat, i.e. indications of a first-order valence transition, which is in agreement with the overall shape of the anomalies.
\begin{figure}[tbp]
	\begin{center}
		\includegraphics[width=\linewidth]{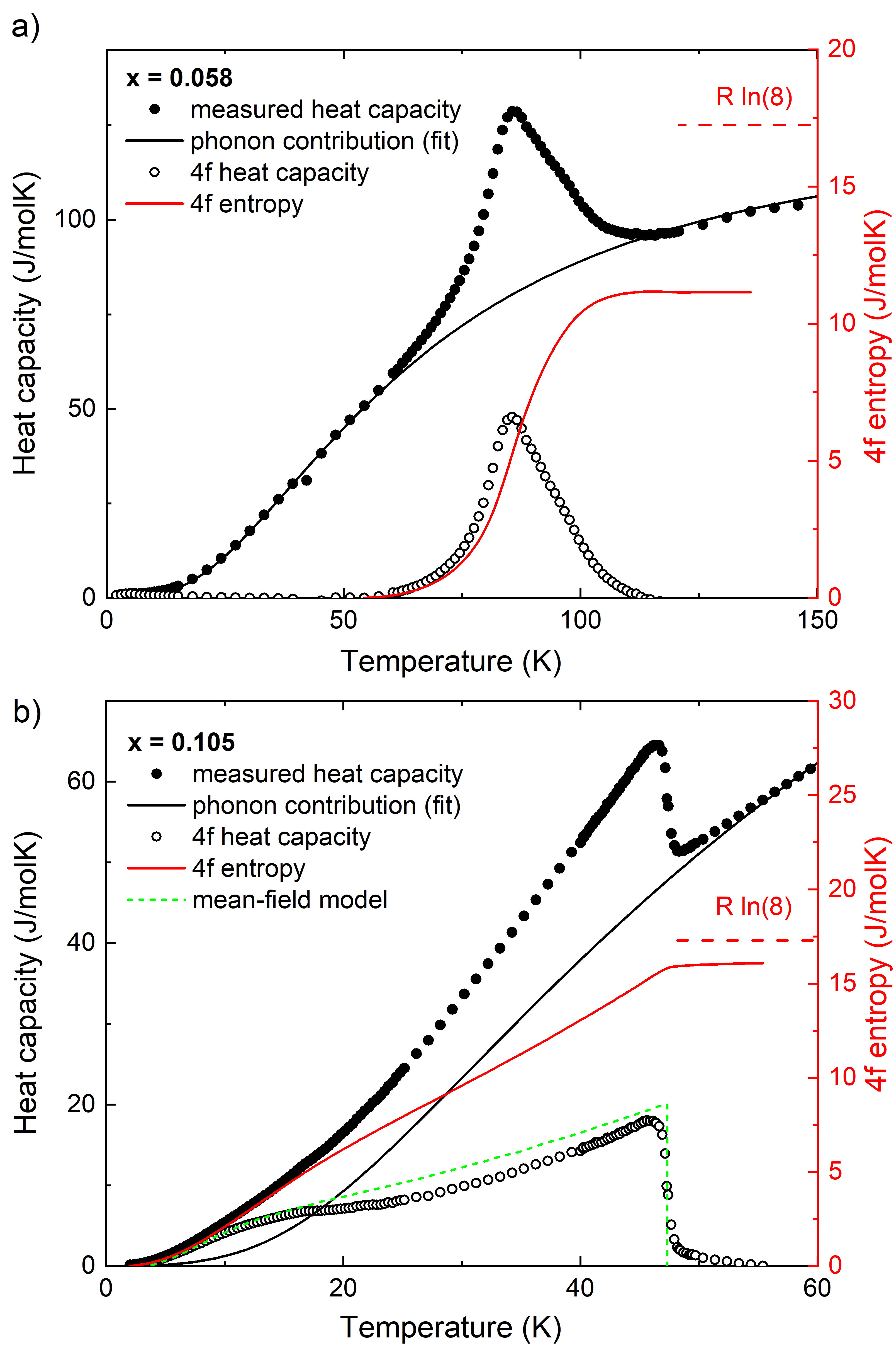}
		\caption[]{Measured heat capacity (solid symbols), phonon background derived from a Debye-fit (black line, see text), $4f$ contribution to the heat capacity (open symbols) and the resulting contribution of $4f$ electrons to the  entropy (red line, right axis) in samples with a) $x=0.058$ (valence fluctuating) and b) $x=0.105$ (antiferromagnetic).  For the AFM system below $T_N$, we included the simulation according to  a mean-field theory from Ref.~\cite{Johnston15} (green dashed curve).
  }
  
		\label{807_HC}
	\end{center}
\end{figure}
\\
To analyze the anomalies of the heat-capacity data for the two different ground states in more detail, the phonon background was determined using an analytic Debye-model function \cite{Anderson19}. We found that using a single Debye temperature did not lead to satisfactory agreement to the measured data, therefore, we allowed for two different Debye temperatures, which led to a significant improvement of the overall fit quality. The resulting Debye fits are shown in Fig.~\ref{807_HC} for the two concentrations $x=0.058$ and $x=0.105$. For the former concentration, we found Debye temperatures of $\Theta_{D1} = (195\pm 10)\,$K and $\Theta_{D2} = (341\pm 5)\,$K, for the latter the high-temperature Debye fit yields $\Theta_{D1} = (162\pm 10)\,$K and $\Theta_{D2} = (346\pm 5)\,$K. The contribution from the conduction electrons (apart from $4f$) to the heat capacity was estimated from the respective contribution to the heat capacity in LaPd$_2$Si$_2$, where a Sommerfeld coefficient of $\gamma=6$\,mJ/molK$^2$ was reported \cite{Besnus91}. Subtracting the phonon and this conduction-electron contribution from the measured data 
 allows us to extract the contribution of $4f$ electrons to the heat capacity, shown as open symbols in Fig.~\ref{807_HC}. It becomes evident, that besides the very different shape of the anomaly at $T'_V$ and $T_N$, the $4f$ contribution is also very different below the characteristic temperatures, resulting from magnonic excitations for the magnetically-ordered compound with $x=0.105$. This contribution is absent for the valence-fluctuating material, without long-range magnetic order. This can be even directly seen in Fig.~\ref{HC}, at e.g. 30\,K, where the heat capacity is much smaller for all valence-fluctuating samples (blue curves) in comparison to the magnetically-ordered samples (red curves). Integrating the $4f$ heat capacity divided by temperature gives the 
 contribution of the $4f$ electrons to the entropy, which is shown as red line in Fig.~\ref{807_HC}. For $x=0.105$ we find an entropy contribution close to $S = \text{R}\cdot\ln(8)$ which corresponds to the expected entropy contribution of a localized magnetic Eu\textsuperscript{2+} ($J=7/2$) moment. In contrast, the contribution of the $4f$ electrons to the entropy found for $x=0.058$ amounts to only $S = \frac{2}{3}\text{R}\cdot\ln(8)$, which accounts for the valence-crossover behavior, i.e., a dynamical admixture of non-magnetic Eu\textsuperscript{3+} ($J=0$) states. For the data of the $x=0.105$ sample, we were able to describe the heat-capacity data using a molecular-field theory for Heisenberg antiferromagnets, developed in Ref.~\cite{Johnston15}. There, the only free parameter to model the heat capacity for a $J=7/2$ system, which orders antiferromagnetially with local moments, is the ordering temperature $T_N$. Using the value $T_N=47$~K for the $x=0.105$ sample, we obtain a theoretical curve, which reproduces the magnetic heat capacity very well (green dashed line in Fig.~\ref{807_HC}b). In particular the shoulder at around $T_N/3$, which is due to the large Zeeman degeneracy of the $J = 7/2$ ground state and frequently observed in magnetically-ordered systems with a $4f^7$ electronic configuration (Eu$^{2+}$ or Gd$^{3+}$), is well described.\\
 
\begin{figure}[tbp]
	\begin{center}
		\includegraphics[width=\linewidth]{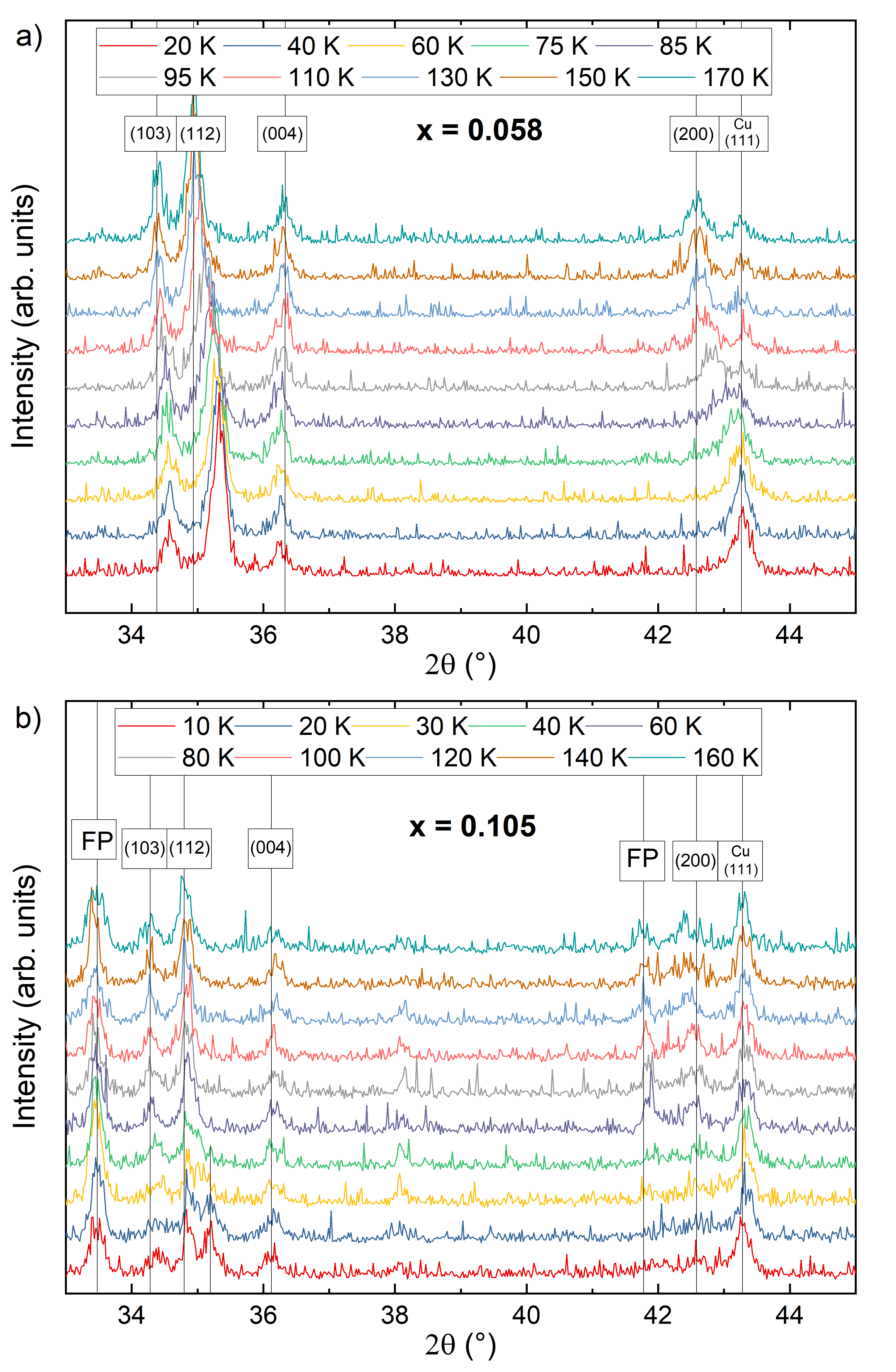}
		\caption[]{Temperature dependent PXRD  data for a) $x=0.058$ (valence fluctuating) and b) $x=0.105$ (antiferromagnetic). Reflections are indexed, FP indicates reflections originating from a foreign phase and Cu marks the reflection from the sample holder.}
		\label{fig_pxrd}
	\end{center}
\end{figure}

\subsection{Lattice effects}
\begin{figure}[tbp]                                 
	\begin{center}                                      
		\includegraphics[width=\linewidth]{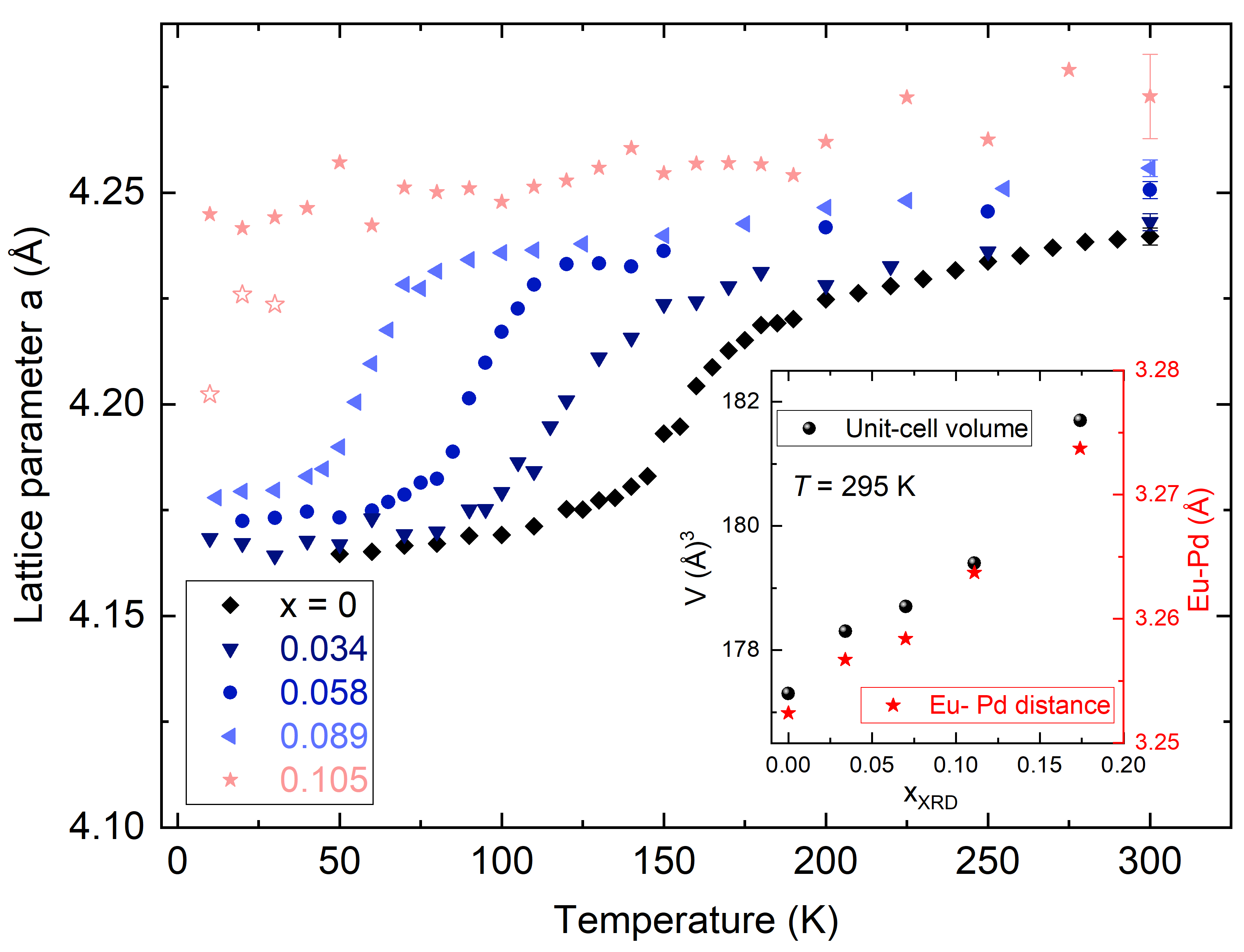}
		\caption[]{Temperature dependence of the lattice parameter $a$ for different Ge-substitution levels in \EPSG. The shown data is obtained by refinement of the temperature dependent PXRD data. 
  The error bars at 300\,K are representative for the temperature dependent powder data. In the inset we present data at $T=295$\,K from the structural characterization of the single-crystalline samples as presented in Table \ref{tab:SCXRD}. The relative change of the unit-cell volume $V$ with $x$ is comparable to the increase of the Eu-Pd distance.}
  
		\label{fig_a}
	\end{center}
\end{figure}

Temperature dependent PXRD data were collected on powdered crystals between 10\,K and 300\,K for the different Ge concentrations with $x\leq 0.105$. The lattice parameters $a$ and $c$ were refined using the established tetragonal unit cell of the ThCr$_2$Si$_2$ structure type (I4/mmm) for $x = 0$. In Fig.~\ref{fig_pxrd}, the temperature-dependent datasets were shown for a) $x=0.058$ and b) $x=0.105$. In the datasets of samples with $x\le0.089$, reflections connected to the $a$ direction of the crystal (most prominently visible for the (112) and (200) reflections) undergo a large shift around the respective characteristic temperature of the valence crossover - so the crystal contracts significantly within the tetragonal plane. For samples with $x=0.105$, no such pronounced shift can be observed. In the dataset shown in Fig.~\ref{fig_pxrd}b, the reflections connected to the $a$ direction remain roughly at the same angle down to 10\,K. However, for the (112) peak, we observe that a sizeable portion of the reflection is shifting out below 50\,K. This suggests, that due to slightly different germanium inclusion levels, some parts of this specific sample is valence fluctuating, leading to a shift of the (112) reflection, whereas the resuming part of this sample shows no strong shift below 40\,K. This observation underlines that for $x=0.105$ germanium incorporation, the system is at the brink of the occurrence of long-range magnetic order. The results of the temperature-dependent PXRD data are summarized in Fig.~\ref{fig_a}, where the $a$ lattice parameter from PXRD is shown as a function of temperature. The data for $x\leq 0.089$ (blue data in Fig.~\ref{fig_a}) all reveal a pronounced anomaly, i.e., a shrinkage of the $a$ lattice parameter upon cooling by about 1.8\%, which agrees well with previous thermal expansion measurements \cite{Wolf22}. The inflection point of the $a$ parameter anomaly coincides well with the inflection point of the magnetic susceptibility (see Fig.~\ref{VSM}) and the maximum of the heat-capacity data (see Fig.~\ref{HC}), tying the lattice anomaly closely to the electronic transition. For the $x=0.105$ sample, we show the lattice parameters below 50~K for both parts of the sample in Fig.~\ref{fig_a}. The open stars reflect the part of the sample which undergoes a valence transition and the closed stars present the reflections which remain at a constant  $2\Theta$ value.  In contrast, overall no strong change is observed for the $c$ lattice parameter within the experimental resolution (not shown), similar to what was observed for $x=0$ \cite{Sampathkumaran81a,Kliemt22}. 

A more thorough structural characterization was done on single-crystalline samples at 295\,K (Table~\ref{tab:SCXRD}). We observe a pronounced increase of the volume of the unit cell with $x$ as shown in the inset of Fig.~\ref{fig_a}. The increase is not perfectly linear, but seem to be larger for small $x$ and significantly stronger towards the point with largest $x$. A similar trend with $x$ was observed for the Eu-Pd bond length, shown as red stars in the inset of Fig.~\ref{fig_a} and the Eu-Si(Ge) bond length (not shown in Fig.~\ref{fig_a}, but given in Table~\ref{tab:SCXRD}). This structural evolution is strongly connected to the observed magnetic ground state in this series. The large increase of the unit-cell volume with increasing $x$ of order 2.5\,\%, leads to a rapid stabilization of the divalent Eu-configuration. In recent \textit{ab-initio} density functional theory (DFT) calculations it was shown that the change of the electronic structure of \EPS\, is intimately related to changes of the Eu-Pd and Eu-Si bond lengths \cite{Song23}. For the antiferromagnetically-ordered system we observe Eu-Pd and Eu-Si(Ge) bond distances of 3.2737(9)\,\AA\, and 3.2611(19)\,\AA, respectively. In comparison to the structural data of other EuT$_2$(Si,Ge)$_2$ systems \cite{Song23}, this would correspond to a divalent ground state of europium. In addition, the temperature evolution of the lattice parameters for $x=0$ are in agreement with the lattice parameters calculated by DFT for the different valence states. When going from Eu$^{2.23+}$ at high temperatures to Eu$^{2.75+}$ at low temperatures \cite{Mimura11}, DFT predicts for \EPS\, a relative change of the $a$ lattice parameters of about 2.5\%, whereas the change for $c$ is significantly smaller and amounts to only 0.26\% \cite{Song23}, which is in good agreement with the observed structural data.

\section{Phase diagram}
\begin{figure}[tbp]                                 
	\begin{center}                                      
		\includegraphics[width=\linewidth]{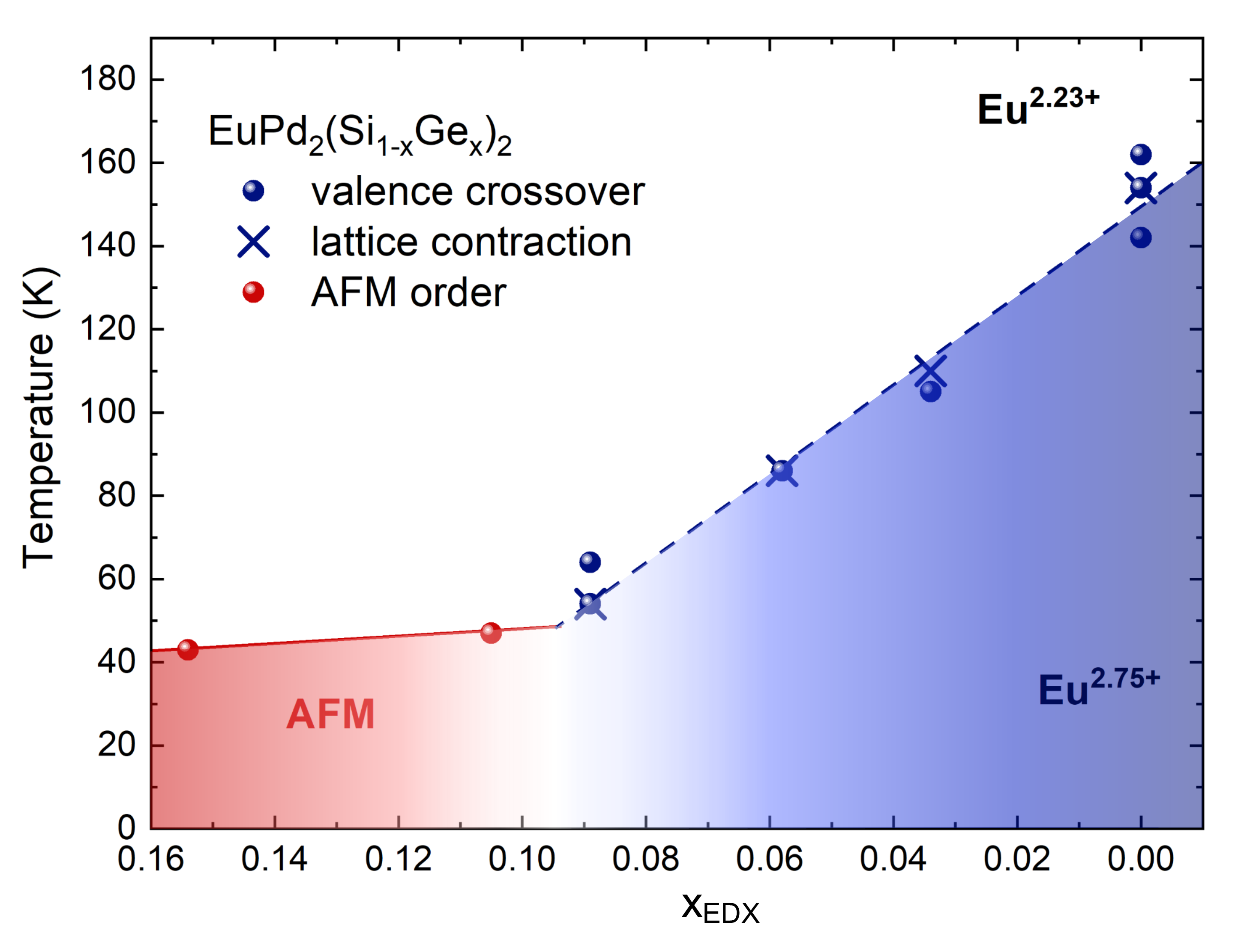}      
		\caption[]{Temperature-substitution phase diagram of \EPSG, denoting the valence-fluctuating regime in blue and the AFM order in red. The red and blue lines are guides to the eye for the AFM phase transition and the valence crossover respectively. Note that the abscissa shows decreasing $x$ values, to be comparable to a positive pressure axis. The high- and low-temperature values for the Eu-valence, Eu$^{2.23+}$ and Eu$^{2.75+}$ were taken from literature data for $x=0$ \cite{Mimura11}.}  
		\label{PD}                                    
	\end{center}
\end{figure}

Bringing all the findings together allows for drawing a concentration-temperature phase diagram for the \EPSG\, system with $0\leq x \leq 0.15$ which is shown in Fig.~\ref{PD}. For low substitution levels $x<0.09$ (blue area), the valence crossover of \EPS\, is maintained, but strongly suppressed down to $\approx 60$\,K. In this region, the thermodynamic signatures of the valence crossover are strongly tied to a pronounced decrease of the in-plane lattice parameter (crosses in Fig.~\ref{PD}). For higher substitution levels $x>0.10$ (red area), the ground state of the system changed to long-range AFM order below $\approx 50$\,K. From these findings it is evident that the critical concentration $x_c$, where valence fluctuations are suppressed and AFM order sets in, is slightly below $x=0.105$. For the next lower concentration level $x = 0.089$ we clearly observe the valence crossover as function of temperature. Therefore, the critical concentration is somewhere in between and we determined it from the presented data as $x_c = 0.10(1)$. 

Ge-substitution of \EPS\, can be seen mainly as chemically induced negative pressure, since Si and Ge are isoelectronically and the volume of the room-temperature unit cell for \EPG\, is about 9\% larger compared to \EPS\, (see Table~\ref{GEincorp}). Assuming a bulk modulus for \EPG\, of the order of $K\approx 80$\,GPa \cite{Onuki20} we can estimate the pressure to reach the volume of \EPS\, when pressurizing \EPG, using $\Delta p = -K\Delta V/V=7$\,GPa. Therefore, the substitution level $x_c$ at which valence-fluctuating behavior changes into long-range magnetic order would correspond to a negative pressure scale applied to \EPS\, of about 0.7\,GPa. The proposition derived from the general pressure-temperature phase diagram of Eu-based materials \cite{Onuki17}, that Eu$^{2+}$ systems undergoes a first-order valence transition under pressure followed by a critical endpoint at finite temperatures, is not precisely seen in the  substitution series studied here. For none of the samples the valence transition showed the characteristics of a first-order phase transition.  Instead, the valence transition remains a rather broad crossover, until it is replaced by the  sharp AFM phase transition for increasing $x$. This is in contrast to the observations made, e.g., in the  series EuNi$_2$(Si$_{1-x}$Ge$_x)_2$ \cite{Wada97} or Eu(Rh$_{1-x}$Ir$_x)_2$Si$_2$ \cite{Seiro11}, where clear indications of first-order valence transitions were observed. 

On the other hand, there exist several Eu-based systems showing a different type of general phase diagram without a first-order valence transition. One example is the series EuCu$_2$(Ge$_{1-x}$Si$_x)_2$, where the occurrence of the valence crossover is observed in direct proximity to the AFM phase at around $x=0.65$ \cite{Fukuda03, Hossain04}.  In addition, resistivity measurements under pressure on EuCu$_2$Ge$_2$ indicate that the AFM transition suddenly drops to zero at a critical pressure of 6.2\,GPa and the authors suggest the existence of a quantum critical point of the valence transition from a nearly divalent state to that with trivalent weight \cite{Gouchi20}. This is corroborated by an increased effective mass and a linear-in-$T$ resistivity around the critical pressure \cite{Gouchi20}.
The measurements presented here suggest that the \EPSG\, series might follow a similar scenario, as we do observe a similar abrupt change from an AFM transition towards the valence-crossover regime. Measurements under He-gas pressure, which will be presented elsewhere \cite{Wolf23}, study this region in more detail and reveal that a critical endpoint at finite temperatures emerges directly out of the antiferromagnetically-ordered state. Further measurements are needed to evaluate this exciting interplay between fluctuating charge-, spin-, and lattice degrees of freedom at around $x=0.1$ in this series. 

\section{Summary}
In conclusion, we have presented the successful single-crystal growth by using the Czochralski technique for a series of \EPSG\, samples with $x_{\text{EDX}}\leq 0.15$. The germanium concentration incorporated into the crystals was found to be significantly smaller compared to the initial concentration from which the crystal growth was started. We found that the valence crossover, established in \EPS\, at $T'_V\approx 160$\,K, can be strongly suppressed with increasing Ge-concentrations down to temperatures of about 60\,K. Remarkably, the character of the valence transition remains crossover-like for all measured samples. At the critical concentration $x_{c}=0.10(1)$ the system changes its magnetic ground state abruptly from valence fluctuating to long-range antiferromagnetically ordered. We observe a sizeable magnetic anisotropy in the ordered state with an easy magnetic plane perpendicular to the tetragonal $c$ direction. For the samples with $x=0.105$ very close to the critical concentration, the magnetic entropy involved in the magnetic transition is close to $\text{R}\cdot\ln(8)$, supporting that in these samples valence fluctuations are practically non-existent. This is reflected also in the temperature-dependence of the lattice parameters for the magnetically-ordered samples, which do not show an additional contraction due to valence fluctuations. In contrast, all valence fluctuating samples show a large continuous change of the the lattice parameter $a$ with temperature of order 1.8\% when going through the valence transition. This corroborates the strong coupling between electronic and lattice degrees of freedom in this series and the crystals with $x=0.105$ are well suited to study their interplay under pressure.

\vspace{5mm}
\section*{Acknowledgments}
We thank K.-D. Luther, T. F\"orster and F. Ritter for their valuable technical support. Discussions with Dominik Hezel and Christoph Geibel are highly appreciated. We acknowledge funding by the Deutsche Forschungsgemeinschaft (DFG, German Research Foundation) via the TRR 288 (422213477, projects A01, A03, and B03).

\bibliographystyle{apsrev}

\end{document}